# 200 mm Wafer-Scale Monolithic 3D Integration of Atomic Layer-Deposited Oxide Semiconductors

Chang Niu[1,2,†], Linjia Long[1,2,†], Luqi Zheng[1,2], Shuting Du[1,2], Jian-Yu Lin[1,2], Kisoo Nam[1,2], Zehao Lin[1,2], Chang Liu[3], Juanjuan Lu[3], Haiyan Wang[3], Haitong Li[1,2] and Peide D. Ye[1,2,*]

[1]*Elmore Family School of Electrical and Computer Engineering, Purdue University, West Lafayette, IN 47907, United States.*

[2]*Birck Nanotechnology Center, Purdue University, West Lafayette, IN 47907, United States.*

[3]*School of Materials Science and Engineering, Purdue University, West Lafayette, Indiana 47907, United States.*

†These authors contributed equally to this work: Chang Niu, Linjia Long

*Correspondence and requests for materials should be addressed to P. D. Y. (yep@purdue.edu)

**Abstract**

**Monolithic 3D (M3D) integration offers a pathway to overcome the scaling limits of conventional silicon complementary metal-oxide-semiconductor (CMOS) technology by extending dense vertical stacking of multifunctional logic and memory devices. Here, we demonstrate wafer-scale M3D integration of three tiers of atomic-layer-deposited (ALD) indium oxide ($InO_x$)-based devices (>100,000 fabricated), including ferroelectric, enhancement-mode, and depletion-mode field-effect transistors, on 200 mm silicon wafers. We achieve threshold voltage standard deviation as low as 0.04 V, average electron mobility up to 91.6 $cm^2V^{-1}s^{-1}$, and fully functional cross-tier circuits. A four-tier 3D computing-in-memory (CIM) accelerator targeting large language model workloads is developed using a custom $InO_x$ process design kit, delivering 1.4× to 2.9× speedup and comparable energy-delay product improvements over 2D baselines. These results establish ALD $InO_x$ M3D integration as a scalable and CMOS-compatible platform for next-generation artificial intelligence hardware and advanced electronics.**

Heterogeneous integration of novel materials and devices is crucial for sustaining transistor scaling in the modern semiconductor industry and for advancing the CMOS+X paradigm with expanded functionalities.[1–5] This integration can mostly be achieved through wafer bonding or other transfer techniques; however, these methods are often limited by the feature size, cost, and success rate. An alternative approach is monolithic 3D (M3D) integration[6–10], which requires back-end-of-line (BEOL)-compatible processes.[11–15] To preserve the functionality of underlying layers, the thermal budget must remain below 400 °C.[16–19] This poses a challenge, as the direct growth or monolithic integration of many novel materials—such as III-V compounds or Ge—typically requires high-temperature epitaxial processes. Atomic layer deposited (ALD) amorphous oxide semiconductors[20–24] offer a promising solution for M3D integration. Nanometer-thin $InO_x$ layers can be uniformly and conformally deposited on various substrates and complex structures using a low-temperature process (~225 °C), making them fully BEOL-compatible. The ALD $InO_x$-based thin-film transistors (TFTs) shows excellent performance, including ultra-high on-current[25], high bias stability[26], low off-current[27], and ultra-low contact resistance[28,29].

In addition to their potential for logic applications, integrating ferroelectric materials with $InO_x$ enables ferroelectric field-effect transistors (Fe-FETs) to exhibit promising characteristics such as a wide memory window (MW), strong retention, and high endurance.[30] With the rising demand for computational power, emerging paradigms such as computing-in-memory and neuromorphic computing require compact integration of multiple functional transistors. M3D integration of vertically stacked, multi-tier architectures combining nonvolatile memory technologies like Fe-FETs with logic devices offers a highly promising route toward energy-efficient, high-performance computing systems. Furthermore, scaling up to 200 mm or 300 mm silicon wafers represents a critical milestone in evaluating this ALD oxide semiconductor technique as a viable option for future manufacturing, alongside other novel channel materials.[16,31–33] For large-scale production, it is essential to evaluate device performance, yield, and uniformity. Statistical analysis of wafer-scale transistors is crucial for assessing both material and process quality, bringing ALD oxide semiconductors one step closer to real-world applications.

In this work, we fabricated and systematically characterized 200 mm wafer-scale, monolithically integrated, multi-tier ALD $InO_x$-based Fe-FETs and FETs, achieving high uniformity, yield, and device performance. Statistical analysis was conducted across a three-tier stack comprising Fe-FETs, enhancement-mode (E-mode) transistors, and depletion-mode (D-mode) transistors. Furthermore, by interconnecting multifunctional devices across different tiers, we demonstrate multi-level ferroelectric nonvolatile memory (MLFe-NVMs), inverters, and embedded dynamic random-access memory (eDRAMs). This oxide semiconductor-based M3D platform is further validated through a computing-in-memory (CIM) accelerator tailored for large language models (LLMs), demonstrating substantial gains in performance and energy efficiency. These results highlight ALD $InO_x$ as a scalable and versatile foundation for next-generation 3D integrated semiconductor systems.

## Atomic layer deposited oxide semiconductors on 200 mm wafers

The schematic structure of M3D integrated multi-tier devices based on ALD $InO_x$ is shown in Figure 1. By leveraging the vertical (out-of-plane) dimension, this architecture enables further semiconductor scaling and heterogeneous integration of diverse functionalities. A 200 mm Si wafer incorporating three vertically stacked tiers of ALD $InO_x$ devices was fabricated, comprising over 100,000 devices across 52 dies. The BEOL-compatible fabrication process, detailed in Method, was used to sequentially integrate Tier 1 Fe-FETs, Tier 2 E-mode FETs, and Tier 3 D-mode FETs. Ferroelectric $HfZrO_2$ (HZO) was employed as gate dielectric for Fe-FETs at Tier 1. High-angle annular dark-field scanning transmission electron microscopy (HAADF-STEM) image and energy dispersive x-ray spectroscopy (EDS) elemental mappings at the gate metal region (Fig. 1d), clearly shows the vertically stacked device layers, including three $InO_x$ channels, ferroelectric HZO, $Al_2O_3$ isolation, and $HfO_2$ dielectric and capping layers. A line-scan along the out-of-plane direction (Fig. S1) confirms the nanometer-scale uniformity of the individual layers and the presence of sharp, well-defined interfaces.

The transfer and output characteristics of short-channel ($L_{ch}$ = 100 nm) and long-channel ($L_{ch}$ = 1 μm) ALD $InO_x$ FETs are shown in Fig. 2. The devices exhibit excellent transistor behavior, including negligible hysteresis, an on/off current ratio exceeding 10 orders of magnitude, a low subthreshold swing (SS = 70 mV/dec), and good saturation in long-channel devices. Short-channel devices demonstrate high drive current, reaching 1000 μA/μm at $L_{ch}$ = 100 nm. The channel length scaling behavior (Figs. 2c and 2d) indicates consistent on-conductance and minor $V_{th}$ roll-off, with small device variation. To comprehensively evaluate device performance and uniformity across wafers, five 200 mm $InO_x$ test wafers were fabricated, each incorporating variations in film thickness and gate dielectric configurations (as shown in Fig. S2). A statistical analysis of over 700 devices per condition at $L_{ch}$ = 5 μm across 200 mm wafers is presented in Fig. 2e, highlighting threshold voltage dependence on $InO_x$ film thickness due to surface electron accumulation. Thicker films exhibit higher $V_{th}$ variability (standard deviation $\sigma_{Vth}$ = 0.18 V for $T_{ch}$ = 3.5 nm), while thinner films improve uniformity ($\sigma_{Vth}$ = 0.04 V for $T_{ch}$ = 1.8 nm), attributed to reduced thickness fluctuation across the wafer (Fig. S3). Thermal treatment provides an effective tuning knob: oxygen annealing is shown to enhance device performance and shift $V_{th}$ (Fig. 2f). Extracted parameters ($V_{th}$, SS, and mobility) across different $InO_x$ thicknesses and annealing conditions are summarized in Fig. 2g. Devices annealed at 300 °C demonstrate improved performance, achieving an average field-effect mobility of 91.6 $cm^2V^{-1}s^{-1}$. The 200 mm wafer-scale ALD $InO_x$ FETs exhibit excellent uniformity, as evidenced by the consistent output characteristics of 137 devices shown in Fig. S4, demonstrating their suitability for large-scale fabrication in semiconductor technology. Building upon these optimizations, wafer-scale ALD $InO_x$ FE-FETs were fabricated by incorporating ferroelectric HZO directly with $InO_x$. The high thermal stability and appropriate interfacial strain provided by $InO_x$ facilitate ferroelectric crystallization in HZO, enabling interfacial-layer-free Fe-FETs. Polarization-voltage (P-V) characterizations of the Fe-FET gate stacks, together with a remanent polarization ($P_r$) map extracted from metal-ferroelectric-semiconductor-metal (Pt-HZO-$InO_x$-Ni) capacitors across 52 dies, confirm uniform ferroelectric performance across the 200 mm wafer (Figs. 2h and 2i).

With excellent device performance and low variability, along with the ability to tune threshold voltage and enhance device stability through annealing, film thickness control, and surface capping, we are now positioned to implement vertical multi-tier integration of ALD $InO_x$-based M3D devices on 200 mm wafers. As shown in Fig. 3, The M3D integration begins with Fe-FETs in Tier 1. The transfer characterization of over 300 Fe-FETs measured on the 200 mm multi-tier wafer under forward and reverse sweeps, reveals a large MW and consistent performance. The devices exhibit a clear distinction between memory states '0' and '1', with no overlap in their distributions. Histograms of the threshold voltages for forward sweeps (average $V_{th}$ = 0.76 V), reverse sweeps (average $V_{th}$ = -1.42 V), and the memory window (average MW = 2.18 V, standard deviation $\sigma$ = 0.24 V), demonstrate tight distributions. The color mapping of the MW extracted from 52 dies across the 200 mm wafer, confirms excellent wafer-scale uniformity. The second-tier ALD $InO_x$ E-mode FETs were fabricated directly on top of the Tier 1 Fe-FETs using BEOL-compatible processes, separated by a 17 nm isolation dielectric layer. The transfer characteristics of over 300 devices measured at different drain voltages ($V_{ds}$ = 0.05 V and 1 V) presented in both linear and logarithmic scales, are shown in Fig. 3d. The histograms in Fig. 3e reveal a tight threshold voltage distribution with minimal variation (average $V_{th}$

= 0.09 V, standard deviation $\sigma$ = 0.11 V), high electron mobility (average $\mu$ = 63.5 $cm^2V^{-1}s^{-1}$), and low subthreshold swing (average SS = 0.124 V/dec). A $V_{th}$ map across the 200 mm wafer confirms the excellent yield and uniform electrical characteristics of the Tier 2 E-mode FETs. The third-tier ALD $InO_x$ D-mode FETs were fabricated using a similar BEOL-compatible process, with a thicker $InO_x$ layer to shift the threshold voltage. Electrical characterization and the corresponding statistical distributions of the Tier 3 D-mode FETs across the 200 mm wafer are presented in Figs. 3g-3i. The increased film thickness resulted in enhanced performance, including a higher electron mobility of 82.0 $cm^2V^{-1}s^{-1}$. Compared to Tier 2 devices, the D-mode FETs also exhibited reduced variation, with an average $V_{th}$ = -1.18 V and a standard deviation $\sigma$ = 0.07 V. The devices fabricated on the top tiers exhibit no performance degradation compared to those on the lower tiers, indicating a robust and reliable integration process. Notably, the vertical stacking of multi-tier devices is not limited to three tiers; additional layers can be integrated through repeated processing steps. The thermal stability of the $InO_x$ BEOL material platform is further supported by the non-degraded electrical performance of Fe-FETs, E-mode FETs, D-mode FETs, and integrated circuits during capping and post-fabrication annealing, including preserved/improved transfer characteristics, SS, $V_t$, $\mu$, and bias stability as shown in Figs. S5-S10. In this work, we successfully demonstrated three-tier M3D integration on 200 mm wafers, confirming that lower-tier devices remain fully functional after the fabrication of upper tiers. This validates the scalability of the approach and its potential for further vertical expansion.

Benchmarking against recent BEOL-compatible channel platforms[34-54] (Table S1 and Fig. S11) shows that our $InO_x$ technology occupies a particularly favorable device-level operating regime. In addition to high mobility, the platform combines wafer-scale process compatibility, low-temperature integration, and suitability for continued multi-tier stacking, distinguishing it from alternative oxide and 2D material approaches. These results indicate that $InO_x$ provides not only competitive transistor performance, but also a more practical foundation for dense monolithic 3D integration.

## Cross tier integrated circuits and memory

Compared to traditional planar circuit architectures, M3D vertically stacked multi-tier logic and memory structures offer significant advantages in area efficiency and design flexibility, enabling integration of multiple functionalities within a compact footprint.[5] In our work, distinct device types are vertically integrated across three tiers: Fe-FETs in Tier 1, E-mode FETs in Tier 2, and D-mode FETs in Tier 3, as illustrated in Fig. 4. This multi-tier configuration enables the realization of multifunctional logic and memory architecture. Specifically, E-mode FETs in Tier 2 are used as access transistors for the Tier 1 Fe-FETs to construct multi-level ferroelectric non-volatile memories (MLFe-NVMs). As shown in Fig. S12, multi-level operation is achieved by storing different voltages at the storage node ($V_{SN}$), facilitated by the ultra-low leakage current of the wide bandgap oxide semiconductors and the stable non-volatile polarization states of the Fe-FETs. This enables the introduction of an additional dimension of data storage. Distinct readout current levels ($I_{rbl}$) corresponding to different storage voltages, modulated by the applied write voltage at different ferroelectric polarization states (down and up), are demonstrated in Figs. 4b and 4c. These distinct current levels allow for expanded multi-level memory operation without overlapping between logic states. We attribute the retention fluctuation mainly to non-ideal parasitic capacitance and layout-dependent coupling in the current proof-of-concept routing scheme, rather than to an intrinsic limitation of the cross-tier structure. By combining n levels of voltage storage at the node and m polarization-induced states in Fe-FET, the MLFe-NVM architecture supports up to $n \times m$ storage levels. This multidimensional storage approach supports high-density, low-power memory with integrated functionality in advanced 3D electronic systems.

Furthermore, cross-tier inverters are demonstrated by integrating Tier 2 E-mode FETs with Tier 3 D-mode FETs. Voltage transfer characteristics and voltage gain of the zero-$V_{gs}$-load cross-tier inverters are shown in Figs. 4d-4g. These inverters exhibit full rail-to-rail swing (from $V_{DD}$ to 0) and a high voltage gain of 23 V/V at $V_{DD}$ = 3 V. The butterfly curve reveals a large noise margin of 65%, confirming robust inverter operation. The voltage transfer characteristics of 52 cross-tier inverters measured across 200 mm wafers, demonstrate excellent electrical uniformity. The distributions of voltage gain and switching threshold (midpoint voltage) show low variation, with a standard deviation $\sigma$ = 0.15 V, as detailed in Fig. S13. Using the same Tier 2 and Tier 3 devices, cross-tier two-transistors zero-capacitor (2T0C) eDRAM cells are implemented, in which the storage node (SN) is formed by connecting the Tier 3 gate and the Tier 2 drain. The key enablers for high-performance 2T0C eDRAMs are the ultra-low off-state current of the access transistor and high bias stability under both positive and negative bias stress (PBS and NBS). The PBS results of the ALD $InO_x$ multi-tier FETs exhibit measurement-limited off-currents and negligible $V_{th}$ shift under stress ($V_S$ = 3 V, equivalent to 5.0 MV/cm for 1000s), even under challenging high-field conditions for high-k dielectrics. The extracted $V_{th}$ shifts for PBS and NBS are only 10.8 mV and -60.3 mV, respectively, as plotted in Figs. 4h and 4i, demonstrating exceptional bias stability. Read currents corresponding to the ‘0’ and ‘1’ memory states at a read voltage of 1 V are shown for the 2T0C eDRAM in Figs. 4j and 4k. A large memory window with on/off current ratio exceeding 8 orders of magnitude is achieved, highlighting the excellent performance of the cross-tier eDRAM. Furthermore, 52 cross-tier eDRAM cells measured across the 200 mm wafer exhibit consistent performance and high uniformity. Together, these results demonstrate fully functional cross-tier logic (inverters) and memory (MLFe-NVMs and 2T0C eDRAMs) integrated on 200 mm wafers, making a significant milestone toward practical monolithic 3D integration based on ALD-grown $InO_x$ semiconductors.

## Indium oxide multi tier computing in memory accelerator

To explore the potential applications and advantages of our 3D multifunctional logic and memory platform, we present a case study featuring a real 3D digital computing-in-memory (CIM) accelerator designed for LLM applications. This design utilizes a custom-developed oxide semiconductor process design kit (PDK) with NMOS-only logic gates based on our platform (Fig. S14). The 3D multi-tier architecture, illustrated in Fig. 5, consists of four vertically stacked tiers integrating analog/mixed-signal circuitry, logic, and both non-volatile and volatile memory, all interconnected via inter-layer vias (ILVs) to realize a true 3D logic-on-memory architecture. The 3D-IC design flow is depicted in Fig. 5b. The entire design is initially partitioned into multiple tiers according to system-level specifications, with each tier represented by a register-transfer level (RTL) digital-top description. The partitioned design is synthesized and hierarchically mapped onto logic gates constructed from our $InO_x$ E-mode and D-mode FETs, generating netlists for each tier. Subsequently, a tier-wise physical design methodology is adopted. Each tier's physical design takes as input both its own netlist and an abstract plus ILV location file from the preceding tier. A standard 2D physical design flow—including floor planning, analog macro placement, standard cell placement, clock tree synthesis, routing, and post-route optimization—is performed for each tier. An example of digital shift & adder implementation with our PDK is shown in Fig. S15. The final output for each tier consists of its GDS file along with an updated abstract and ILV location file, which are passed on to the next tier in the design sequence.

The $InO_x$ multi-tier 3D CIM macro is shown in Fig. 5c. Tier-1 comprises a 4Mb non-volatile ferroelectric memory bank utilizing Fe-NVM array to store all static weights of the LLM model. The second tier integrates analog/digital peripheral circuits supporting the Fe-NVM array in Tier-1, including word line (WL) drivers for programming and sense amplifiers for data readout. Tier-2 also incorporates digital CIM logic, such as adder trees, shift-and-add logic, local controllers, and register-based input/output buffers for enabling FeFET-based non-volatile computing-in-memory (nvCIM). Tier-3 integrates a 2Mb 2T gain-cell eDRAM macro, comprising both the 2T array and corresponding peripheral circuits, to support dynamic computation in the LLM model. Benefiting from the ultra-long retention time of the 2T cell, the design achieves a refresh-free eDRAM macro implementation. Tier-4 is the top tier of the design, containing digital CIM logic interfacing with Tier-3 eDRAM and a top-level cross-tier controller for orchestrating computation and data movement across all four tiers. The complete design consists of approximately 1.2 million logic gates and is implemented using a custom-developed oxide semiconductor PDK, scaled to the 40 nm technology node. The final merged 3D-IC GDS layout is illustrated in Fig. 5d.

As shown in Fig. 5e, the backbone operation of the Transformer-based LLM (Fig. S16) can be categorized into static and dynamic classes based on the nature of matrix-vector multiplication (MVM). In the multi-head attention mechanism, the Q, K, and V projections are static, as they originate from the pre-trained backbone, making them well-suited for non-volatile nvCIM in Fe-NVM. In contrast, the computation of the attention scores $Q \times K^T$ and $A' \times V$ is dynamic, rendering it more suitable for CIM in dynamic eDRAM. Leveraging the monolithic integration of Fe-NVM and eDRAM in the 3D CIM macro and scaling up our 3D CIM macros, we propose an end-to-end LLM accelerator design, as illustrated in Fig. 5f. This architecture integrates 3D CIM macros, a softmax unit, accumulators, a global buffer, and a network-on-chip (NoC). The LLM accelerator is evaluated using our in-house framework[55], which performs comprehensive model-to-chip mapping by processing detailed model parameters, chip configuration settings, and calibrated M3D technology libraries (Fig. S17). The framework systematically compiles model layers into static and dynamic operations, optimizes operation scheduling and dataflows through exploration of diverse mapping schemes, guided by the architectural characteristics of the 3D CIM macro. Performance metrics of the accelerator—including energy consumption, speedup, and energy-delay product (EDP) improvements—are evaluated for inference workloads across state-of-the-art LLM models such as BERT, GPT-2, and DeiT (as shown in Fig. 5g).[56–59] Compared to a conventional 2D design, the proposed M3D architecture achieves a 1.4× to 2.9× speedup and comparable EDP benefits for GPT-2 and BERT-small models, respectively, demonstrating system-level performance enhancements enabled by the M3D design and technology platform. The 2D baseline is defined as a planar OS-FET implementation under the same footprint constraint and NVM/eDRAM composition ratio, using the same workload mapping strategy and model configurations. This formulation captures the impact of M3D integration in terms of effective resource density and data movement. Benchmarking against state-of-the-art systems (Table S2), including 2D-material-based processors,[60] CNT-based processors,[47] and recent 3D memory-centric architectures,[61,62] highlights the key system-level advance of this work: the combination of enhanced integration capability and broader workload support.

**Conclusions**

M3D integration overcomes the limitations of traditional two-dimensional architectures by enabling the vertical stacking of transistors with diverse functionalities on separate tiers, tightly interconnected to form compact, multifunctional systems. We demonstrated a wafer-scale M3D platform based on ALD-grown $InO_x$ devices, achieving vertically stacked and BEOL-compatible integration of ferroelectric, E-mode, and D-mode FETs on 200 mm silicon wafers. The fabricated devices exhibit excellent electrical performance, high uniformity, and robust cross-tier functionality, enabling the realization of multifunctional logic and memory systems in a compact 3D structure. By leveraging this platform, we implemented a full-stack CIM accelerator tailored for LLM workloads, showcasing the integration of non-volatile and volatile memory with digital and analog logic across four vertically interconnected tiers. System-level evaluations highlight significant performance and energy-efficiency gains over conventional 2D architectures. Further scaling will require systematic design of the three-dimensional stack, tighter control of oxide-semiconductor processing to tailor device characteristics, and careful management of parasitic coupling and heat transport, which cannot be inferred directly from two-dimensional circuits. Larger hardware demonstrations are also needed to validate the projected computing-in-memory performance and energy efficiency.

## Figures

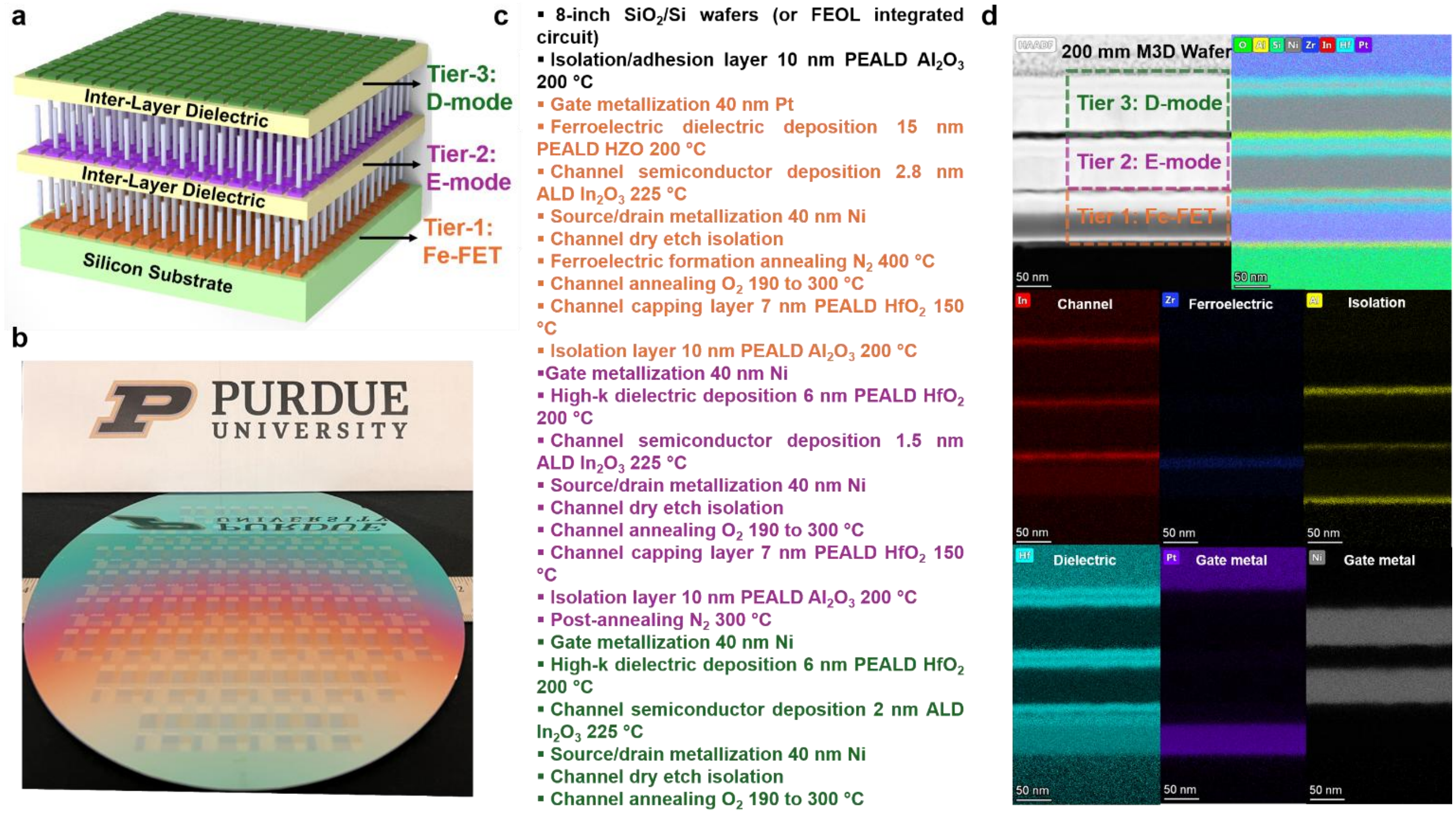

**Fig. 1. Multi-tier ALD $InO_x$ integration on 200 mm wafers. a**, Schematic illustration of the M3D integration of ALD $InO_x$-based devices, including ferroelectric, E-mode, and D-mode FETs across three tiers. **b**, Optical image of a 200 mm silicon wafer with integrated multi-tier ALD $InO_x$ devices. **c**, Fabrication process flow for the wafer-scale M3D integration, showing the sequential stacking of three transistor tiers using BEOL-compatible processes. **d**, HAADF-STEM image and corresponding EDS elemental maps of the device cross-section, confirming the vertically stacked architecture and spatial distribution of the integrated materials.

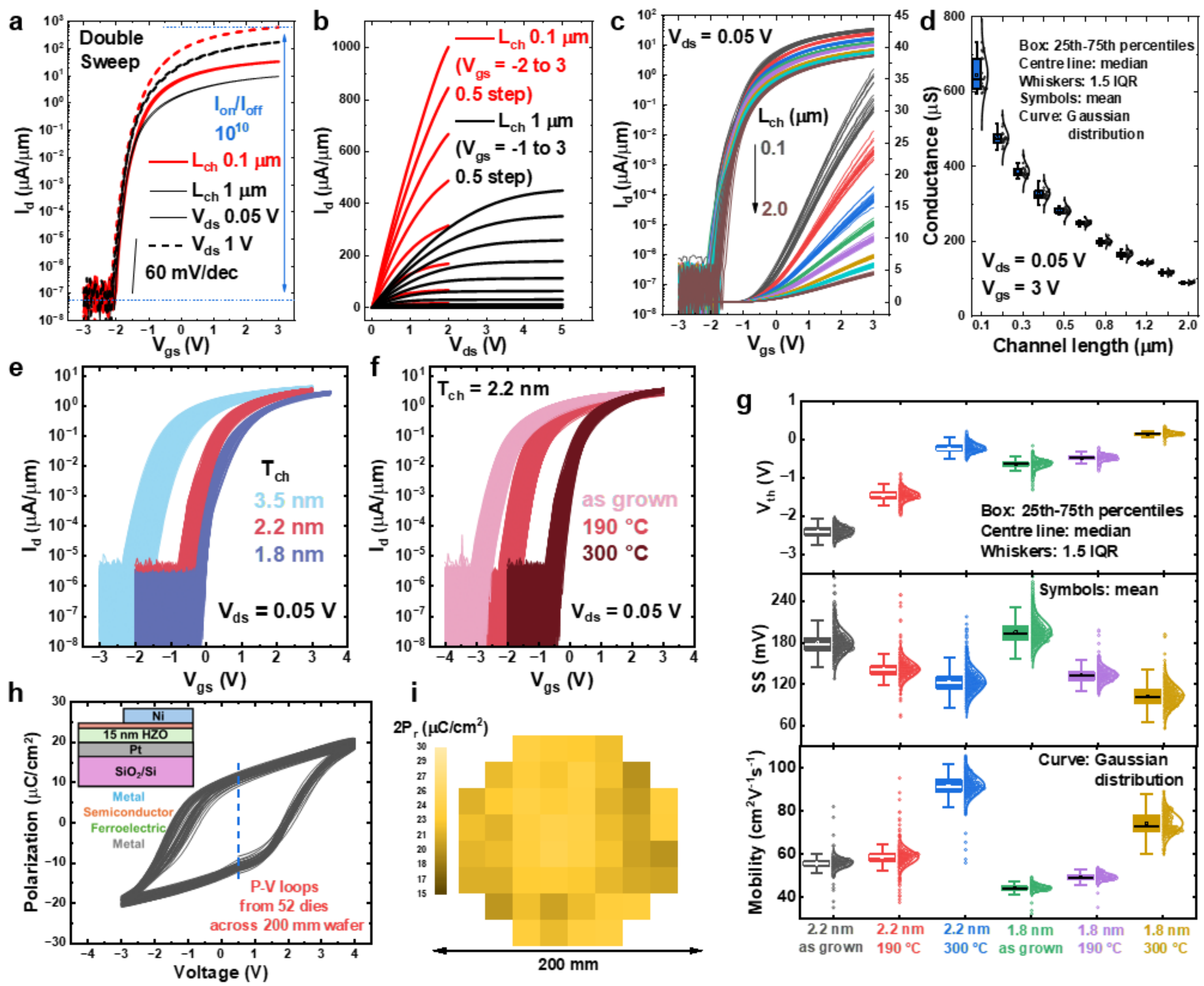

**Fig. 2. Process-dependent electrical performance of ALD $InO_x$ on 200 mm wafers. a-b**, Transfer and output characteristics of $InO_x$ FETs, showing high on-current (1000 µA/µm) in short-channel devices and excellent saturation behavior in long channel devices. **c-d**, Channel length dependence of the transfer characteristics, demonstrating the scalability of $InO_x$ FETs. The sample sizes were n = 19, 19, 21, 20, 19, 20, 18, 16, 20, 20, and 18, respectively. Centre symbols indicate mean values. **e**, Statistical analysis of device characteristics for varying $InO_x$ thicknesses across 200 mm wafers. **f**, Statistical analysis of device characteristics under different annealing temperatures. **g**, Extracted device parameters (threshold voltage, subthreshold slope, and mobility) from 200 mm wafers. The sample sizes were n = 1601, 1606, 1596, 1601, 1600, and 753, respectively. Centre symbols indicate mean values. **h**, Polarization-voltage (P-V) loop of metal-semiconductor-ferroelectric-metal capacitors on 200 mm wafers. **i**, Color map of remanent polarization across a 200 mm wafer, confirming spatial uniformity.

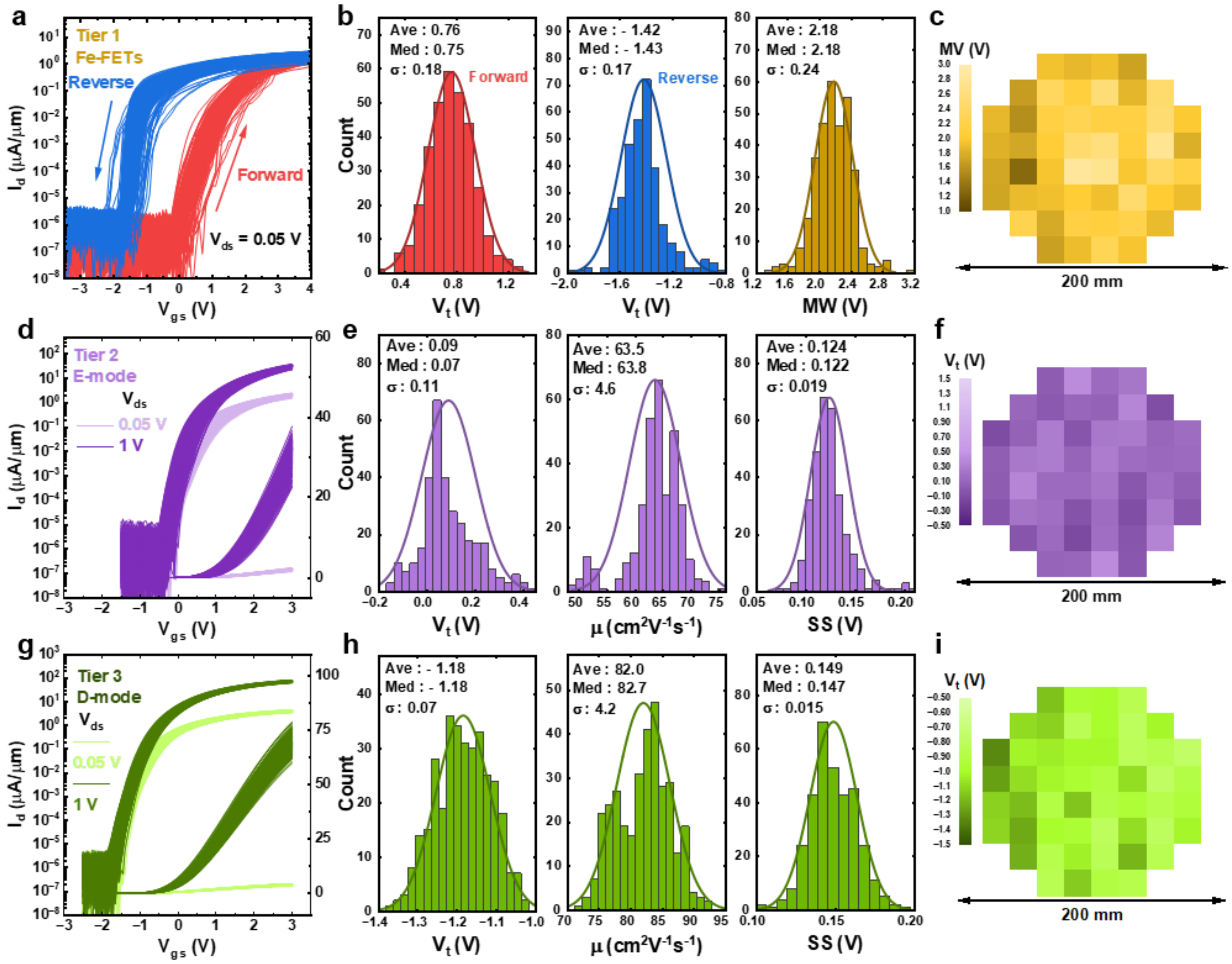

**Fig. 3. Device characterizations of multi-tier ALD $InO_x$ transistors on a 200 mm wafer.** **a-c**, Electrical performance of Tier 1 Fe-FETs: transfer characteristics (**a**), statistical distributions of extracted parameters (**b**), and spatial color map of threshold voltages across the wafer (**c**). **d-f**, Electrical performance of Tier 2 E-mode FETs: transfer characteristics (**d**), extracted device statistics (**e**), and threshold voltage mapping across the wafer (**f**). **g-i**,

Electrical performance of Tier 3 D-mode FETs: transfer characteristics (**g**), extracted device statistics (**h**), and threshold voltage mapping across the wafer (**i**).

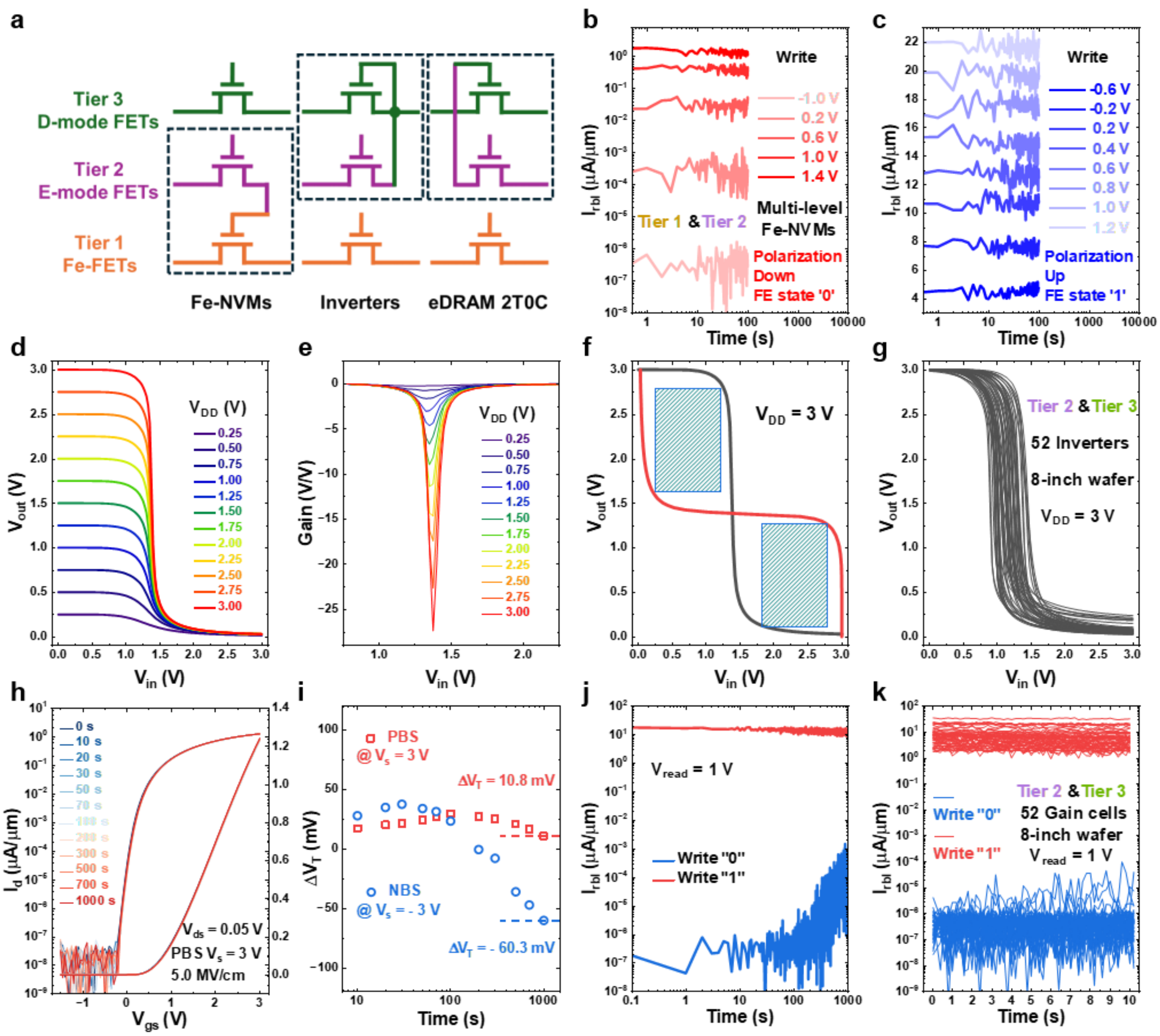

**Fig. 4. Cross-tier multifunctional vertical integrated logic and memory on a 200 mm wafer. a**, Schematic illustration of cross-tier logic and memory integration, including Fe-NVMs, inverters, and 2T0C eDRAM. **b-c**, Multi-level operation of a Fe-NVM cell under opposite ferroelectric polarization states (up and down). **d-g**, Electrical performance of cross-tier inverters: voltage transfer characteristics (**d**), voltage gain (**e**), butterfly curve

showing the noise margins as shaded regions (**f**), and uniform transfer characteristics of 52 inverters across the 200 mm wafer (**g**). **h-k**, Device performance of cross-tier 2T0C eDRAM cells: bias stability under positive (**h**) and negative (**i**) stress conditions, retention characteristics showing large on/off ratio (**j**). and consistent readout current levels across 52 cells measured over the 200 mm wafer (**k**).

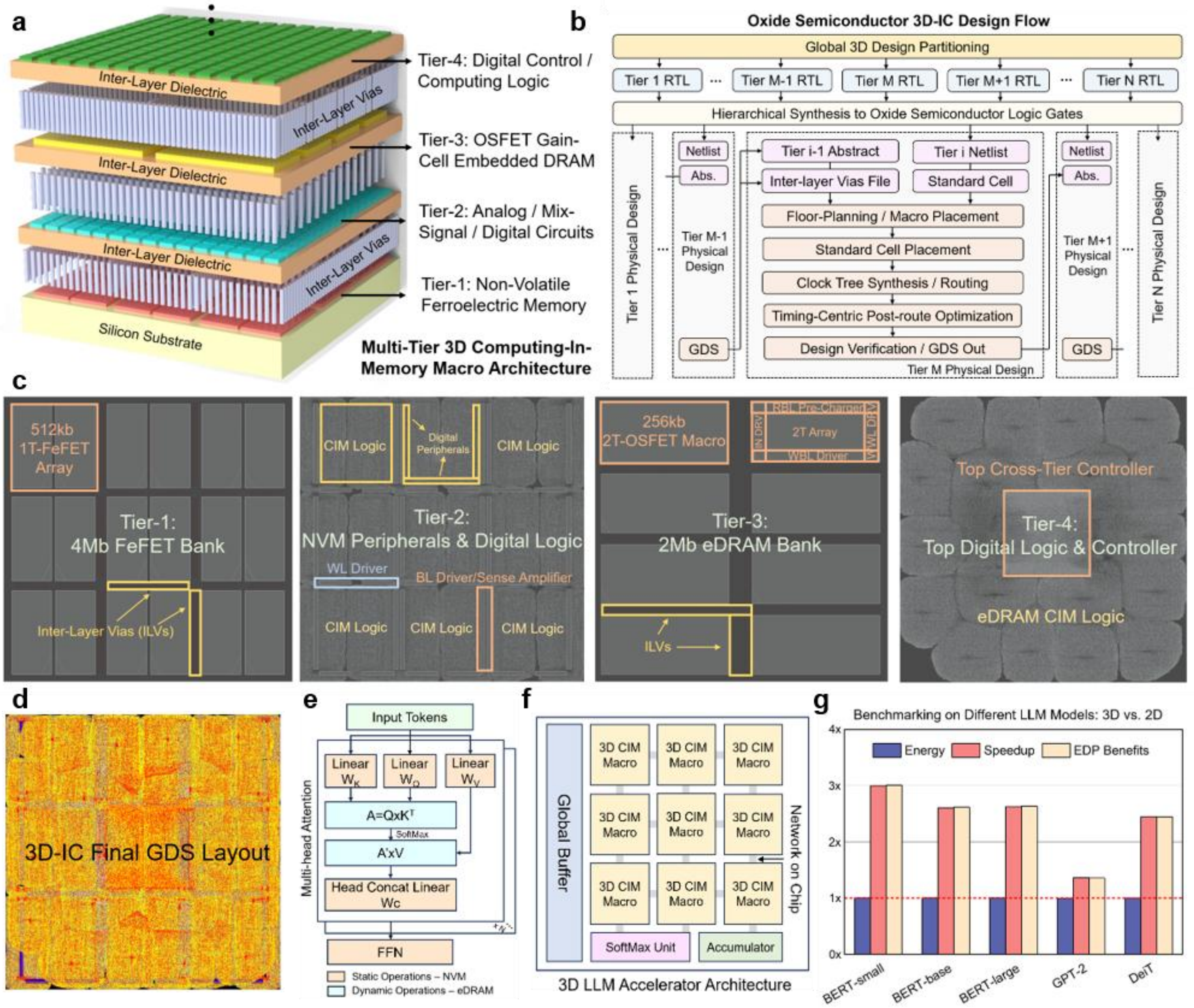

**Fig. 5. $InO_x$-based 3D multi-tier computing-in-memory (CIM) LLM accelerator design and benchmarking.** **a**, Schematic of the proposed multi-tier 3D CIM macro, featuring four tiers with scalability for additional stacking. **b**, Customized 3D-IC design flow from RTL to GDS using PDK based on our $InO_x$-based oxide semiconductor platform.

**c**, Physical implementation of the multi-tier CIM macro, integrating non-volatile and dynamic memory, analog peripherals, digital compute, and control logic across vertical tiers. **d**, Final merged layout of the complete multi-tier 3D CIM macro. **e**, Partitioning of Transformer model computation into static (non-volatile memory) and dynamic (eDRAM-based) operations. **f**, Top 3D LLM accelerator architecture with multiple 3D CIM macros interconnected via network on chip. **g**, System-level benchmarking results showing energy, delay, and energy-delay product comparisons between 3D and 2D architectures across various LLM models.

## Methods

**Atomic-Layer-Deposition (ALD) and device fabrication.** Full fabrication process flow is shown in Fig. 1c. The device fabrication process began with 200 mm Si wafers with 1.5 μm $SiO_2$, followed by a standard cleaning process, including ultrasonic rinsing with toluene, acetone, and isopropyl alcohol to remove possible organic particles and dirty materials. A 10 nm layer of $Al_2O_3$ is deposited using plasma-enhanced atomic layer deposition (PE-ALD) at 200 °C. A standard lift-off process was then applied for the 40 nm Ni or Pt bottom gate for FETs and Fe-FETs. Next, a 6 nm $HfO_2$ dielectric or 15 nm $HfZrO_2$ (HZO) ferroelectric layer was deposited by PE-ALD at 200 °C, using $[(CH_3)_2N]_4Hf$ (TDMAHf) and $[(CH_3)_2N]_4Zr$ (TDMAZr) as Hf and Zr precursors. Subsequently, 1-3 nm $In_2O_3$ was deposited by ALD at 225 °C using $(CH_3)_3In$ (TMIn) and $H_2O$ as In and O precursors. The TMIn precursor was heated to 60 °C to provide sufficient vapor pressure. Film thickness was accurately controlled by the number of ALD cycles. Channel isolation was done by dry etching the $In_2O_3$ layer using Ar plasma. A 40 nm Ni layer was deposited as source/drain contacts. The devices were then annealed at various temperatures (190 to 300 °C) in an $O_2$ atmosphere. For ferroelectric layer activation, a 400 °C $N_2$ annealing was conducted. A 7 nm $HfO_2$ capping layer was deposited by PE-ALD at 150 °C, followed by a 10 nm PE-ALD $Al_2O_3$ insulating layer at 200 °C. Multi-tier structures were realized by repeating the fabrication steps. Three-tier oxide semiconductor Fe-FETs and FETs were successfully demonstrated on 200 mm Si wafers. Notably, the entire fabrication process is compatible with back-end-of-line (BEOL) integration, with a maximum thermal budget of 400 °C. In the present M3D device demonstration, the layout has not yet been optimized to minimize parasitic capacitance and coupling, both of which can be further reduced in future designs.

**Wafer-scale electrical measurements and statistical methods.** The fabricated multi-tier 200 mm wafer was characterized using a FormFactor PAV200 200 mm semi-automated probe station at 25 °C in $N_2$ environment at ambient pressure. The electrical characterization of transistors, inverters, and memories was measured with the Keysight B1500A Semiconductor Device Parameter Analyzer. The polarization vs. voltage loops were measured using Radiant Premier II Ferroelectric Tester. Unless otherwise stated, the characterized devices had a channel length of 5 µm.

The subthreshold swing was extracted from the inverse slope of the transfer characteristic around a normalized drain current of $10^{-4}$ µA/µm. The threshold voltage was extracted using the constant-current method at a normalized drain current of $10^{-3}$ µA/µm. The field-effect mobility was extracted in the linear regime using the maximum-transconductance method. All device parameters were extracted from transfer characteristics measured at $V_{ds}$ = 0.05 V. Threshold voltage shifts in the bias stability tests were determined using the linear-extrapolation method.

Box plots show the 25th–75th percentiles, with centre lines indicating the medians and centre symbols indicating the means. Whiskers extend to the most extreme data points within 1.5 × the interquartile range (IQR), where the IQR is the difference between the 75th and 25th percentiles. For the transfer characteristic overlays in Fig. 2, devices classified as outliers were excluded from the displayed curves to facilitate visualization of process-dependent device behaviour. Gaussian curves were generated using the sample means and standard deviations. Exact sample sizes are provided in the corresponding figure legends.

**High resolution scanning transmission electron microscopy (HR-STEM).** TEM, selected area diffraction, energy dispersive x-ray spectroscopy (EDS) elemental mappings, and HAADF-STEM analysis were performed with an FEI Talos F200X operated at 200 kV. Samples were prepared by focused ion beam (FIB) lift-out using a Helios G4 UX dual-beam system.

**Oxide-semiconductor process design kit (OS-PDK).** A custom oxide-semiconductor process design kit was developed based on the measured electrical characteristics of the InOx enhancement-mode and depletion-mode transistors. Parameterized Verilog-A transistor models were created for device- and circuit-level simulation in Cadence Virtuoso. An NMOS-only standard-cell library was constructed using enhancement-mode transistors as pull-down switching devices and depletion-mode transistors as active loads. The logic functions and transient responses of the standard cells were verified through transistor-level simulations. The OS-PDK includes Liberty-format (LIB) files describing the logic, timing, and power characteristics of the standard cells, library exchange format (LEF) files containing their physical abstracts, technology files (TechLEF) defining the routing layers and inter-layer via (ILV) design rules, and GDS files containing the complete standard-cell layouts.

**Circuit and physical design flow.** The memory arrays and analog or mixed-signal peripheral circuits were designed and simulated in Cadence Virtuoso using the custom Verilog-A transistor models. The digital computing and control modules were described in Verilog HDL and functionally verified using Questa. Logic synthesis was performed using Cadence Genus, where the RTL designs were hierarchically mapped to the custom InOx standard-cell library. The synthesized netlists were then implemented in Cadence Innovus using a tier-by-tier 3D physical design flow. For each tier, the physical design process included floorplanning, analog macro placement, standard-cell placement, clock-tree synthesis, routing, and post-route optimization. In addition to its own synthesized netlist, each tier used the physical abstract and ILV location file generated from the preceding tier as design inputs. After physical implementation, an updated abstract file and ILV location file were exported together with the GDS layout and passed to the subsequent tier. This sequential flow maintained the alignment of the inter-layer vias and ensured consistent cross-tier connectivity. Finally, the layouts of all tiers were vertically aligned and merged to generate the complete 3D-IC GDS database.

**System simulation and benchmarking.** System-level evaluation was performed using our model-to-chip mapping framework. The framework takes the model topology and parameters, chip configuration, and calibrated $InO_x$ technology libraries as inputs. Circuit-level characteristics were obtained from simulations of the memory arrays, peripheral circuits, and standard cells using the custom oxide-semiconductor PDK. For each workload, model-layer computations were partitioned into static and dynamic matrix–vector multiplication operations. These operations were mapped to the FeFET- and eDRAM-based CIM resources, respectively, within the 3D macro. Network-on-chip traffic patterns were simulated using the router microarchitecture and trace-driven traffic injection, and the associated communication costs were evaluated based on physical-design considerations. By integrating the technology-library parameters with layer-wise execution traces, the end-to-end latency, energy consumption, and energy–delay product were estimated. Each inference workload was evaluated using identical model configurations for the 3D architecture and its 2D baseline. The 2D baseline was a planar oxide-semiconductor implementation constrained to the same chip footprint, FeFET/eDRAM composition ratio, and workload-mapping strategy as the 3D design.

**Acknowledgements**

P.D.Y. and H.L. were supported in part by the U.S. National Science Foundation under Award No. 2425498 with industry partners as specified in the Future of Semiconductors (FuSe2) program. P.D.Y. was also supported by Semiconductor Research Corporation Global Research Program and Samsung Electronics, Inc. C. L., J. L. and H.W. acknowledge the support from the U.S. National Science Foundation for the microscopy work (DMR-2016453 and DMREF-2323752). We acknowledge Cadence for Innovus 3D-IC license support.

**Author Contributions**

P.D.Y. supervised the project. C.N. designed the experiments. C.N., L.L., Z.L., and J.-Y. L. fabricated the devices. L.Z., S. D., and H.L. conducted the circuit-level simulation. C.L., J.L. and H.W. carried out the TEM/STEM measurements and image analysis. C.N., L.L., J.-Y.L. and K.N. performed the electrical measurements and analyzed the data. P.D.Y. and C.N. wrote the manuscript and all the authors commented on it.

**Competing Interests**

The authors declare no competing interests.

**Supporting Information**

Additional details for 200 mm ALD $InO_x$ test wafers, statistical distribution of device parameters, statistical analysis of the output characteristics, working principle of MLFe-NVM, ALD $InO_x$ PDK, and circuit design for LLMs can be found in supplementary information.

## Corresponding Author

* Peide D. Ye (E-mail: yep@purdue.edu)